\documentclass[twocolumn,10pt]{wlscirep}
\usepackage{geometry}
\usepackage{lmodern}
\usepackage{float}
\usepackage{tabularx}
\usepackage{chemfig}
\usepackage{chemobabel}
\usepackage{adjustbox}
\usepackage{dirtytalk}
\usepackage{booktabs}
\usepackage[utf8]{inputenc}
\usepackage[version=4]{mhchem} 
\usepackage{multirow}
\usepackage{orcidlink}
\usepackage{amsfonts,amsmath,amssymb}
\usepackage[T1]{fontenc}
\usepackage{widetext}
\usepackage{footnote}
\usepackage{cleveref}
\title{\textbf{Partial ionization cross sections of isobutanol}}
\author{Suriyaprasanth Shanmugasundaram \thanks{\href{mailto:{suriyaprasanth.s@vit.ac.in}}{suriyaprasanth.s@vit.ac.in}}\orcidlink{0000-0002-3939-1446}}
\author{Dhanoj Gupta\thanks{Corresponding author: \href{mailto:dhanoj.gupta@vit.ac.in}{dhanoj.gupta@vit.ac.in}} \orcidlink{0000-0001-6717-8194}}
\affil{Department of Physics, School of Advanced Sciences,
Vellore Institute of Technology, Katpadi,
Vellore - 632014, Tamil Nadu, India.}

\keywords{BEB model, isobutanol, branching ratios}

\begin{document}

%
%

\begin{abstract}
\rm The electron and positron impact partial ionization cross sections (PICS) for isobutanol were calculated using variants of the binary encounter Bethe model (BEB). The modified BEB (mBEB) model and the mass spectrum dependent (MSD) method are used to calculate the PICS of cationic fragments of isobutanol. Also for calculating the positron impact ionization and dissociative ionization cross sections, we have used the BEB-0 model and the BEB-W model which are scaled Wannier threshold laws along with the BEB-A and BEB-B models which obey Jansen's threshold law. The study also compares the PICS calculated from other isomers of butanol presented in the literature. We also presented the synthetic electron impact mass spectrum of isobutanol which is compared with the experimental measurements.
\end{abstract}

\maketitle

\section{Introduction}

The isobutanol is an important biofuel that has been commonly used as an additive in sustainable aviation fuel\cite{su2023biofuels}. Like ethanol, isobutanol can be derived from fermentation of sugars or common agricultural wastes. Isobutanol and ethanol are an important intermediates in the sugar fermentation pathway in the production of jet fuel\cite{su2023biofuels}. The production of liquid fuels such as gasoline (petrol), diesel, and isooctane is derived from isobutanol through one or more chemical pathways; more information on sustainable aviation systems can be obtained from the work of Su-ungkavatin et al.\cite{su2023biofuels} Interest in biofuels has also increased due to the risk associated with recycling and mining Lithium resources worldwide. Many countries have agreed and announced to reduce their net CO$_2$ emission to 45 and 50 \% by 2030. \cite{UN_NetZero_Coalition} This was also part of the sustainable development goals of the United Nations. Thus, the importance of biofuels and their studies is underlined.

In the literature, the review of biofuels by Lopes et al.\cite{lopes2020electronbiofuel} consolidates all studies performed until 2020 in which they have consolidated mass spectrometry, partial ionization cross sections (PICS), total cross sections (TCS), momentum transfer cross sections (MTCS), total ionization cross sections (TCS) and integrated cross sections (ICS) of primary alcohols such as methanol, ethanol, 1-propanol and 1-butanol. The complete experimental study of the electron impact mass spectrum, appearance energies, and PICS measurements for 1-butanol and 2-butanol is performed by Lopes et al.\cite{pires2018electron,ghosh2018electron,amorim2022-2butanol-1,amorim20232-butanol-2}. On the theoretical front, the PICS for the 1-butanol cations was calculated by Goswami et al.\cite{butanol-goswami2022electron} using the mBEB method, and for 2-butanol we calculated the PICS of the cations using the MSD and mBEB method in our previous work\cite{shanmugasundaram2024electron}. The results of these calculations are in reasonable agreement with the experimental PICS data.  The electron impact mass spectrum (EIMS) of isobutanol was measured by Oliveira et al.\cite{oliveira2024ionic} using the HIDEN EPIC 300 quadrupole mass spectrometer with a range of 1 to 300 amu with a mass resolution of 1 amu, coupled with the mass spectrometer interface unit, the energy pulse ion counter probe (EPIC) operated in residual gas analysis mode (RGA). The EIMS was recorded at 70 eV, where 45 cations were detected that contained six new cations presented in their work.

\begin{figure}[hb]
    \centering
    \includegraphics[width=0.25\textwidth]{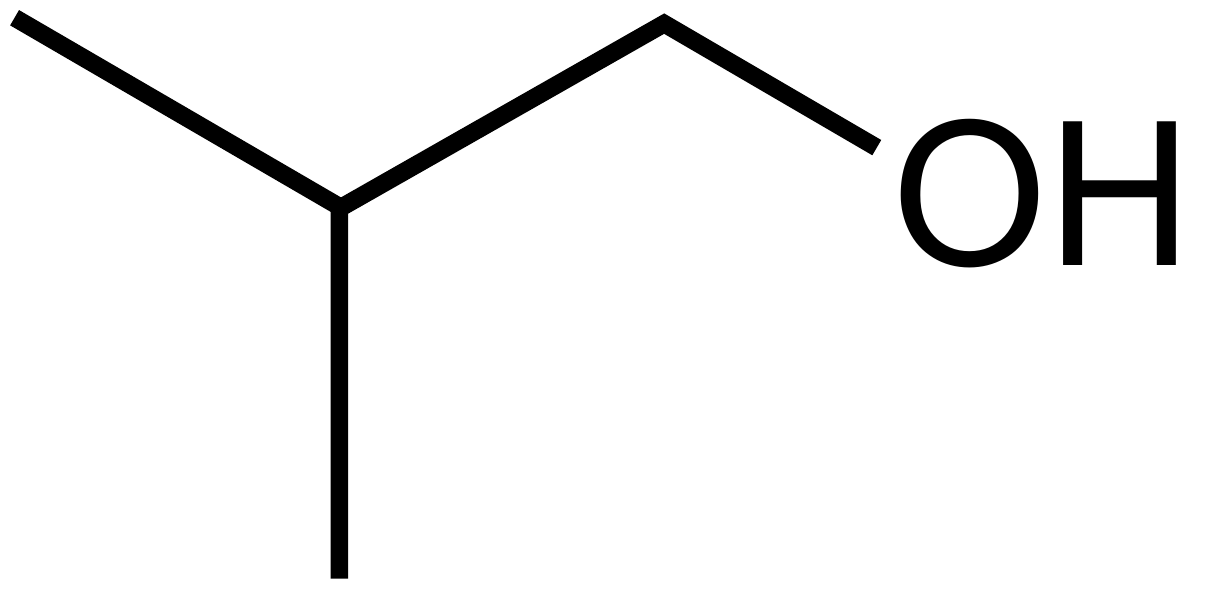}
    \caption{isobutanol ($\mathsf{C_4H_{10}O}$)}
    \label{fig:struct}
\end{figure}

Our aim in this work is to calculate the theoretical PICS for the electron and positron impact for the isobutanol fragments making use of the appearance energies and mass-spectrum data from the experiment. As there is no data available for the PICS of isobutanol, our results are compared with the experimental PICS data of 1-butanol and 2-butanol for similar cationic fragments to draw important conclusions. The MSD, m-BEB, and BEB methods in combination with mass spectrum data are used to obtain the PICS and TICS of isobutanol. The structure of isobutanol is shown in figure \ref{fig:struct}. The structure of the manuscript is as follows. In section \ref{sec:level2} we introduce the BEB model and their variants, section \ref{sec:pics} we show the method to calculate the PICS, and, finally, section \ref{sec:rd} contains results and discussions.

\section{BEB Model}\label{sec:level2} 
The BEB model \cite{kim1994binary} is one of the most common and widely used methods for calculating TICS for atoms, molecules, radicals and ions and is being used here with some modifications to calculate the PICS. The BEB method calculates the electron impact ionization cross section for each orbital, and the sum of the ionization cross section for each orbital gives the total ionization cross section as shown in Eq. \eqref{eq:1} for a target under consideration. Eq. \eqref{eq:2} gives the BEB formula for determining TICS.
\begin{equation}\label{eq:1}
    \sigma_{\rm TICS}(E) = \sum_{i}^{N} \sigma_{i}~(E)
\end{equation}

\begin{equation}\label{eq:2}
 \sigma_i~(E) =\alpha \left [ \frac{ \ln t_i}{2}\left ( 1 - \frac{1}{t_i^{2}}\right) +  \left ( 1 - \frac{1}{t_{i}} \right )- k  \right ]
\end{equation}
The variable $\alpha$ houses the Burgess denominator $(t_i+u_i+1)$, which serves as an $adhoc$ correction associated with the effective kinetic energy experienced by the target electron, and the $(u_i+1)$ term is acceleration due to nuclear attraction.
\begin{equation}
    \alpha = \frac{\rm S_i}{(t_i+u_i+1)}
\end{equation}
Once including $\alpha$, along with the Mott theory for hard collisions and Bethe theory for soft collisions and electron-electron exchange effects, gives us the simplified BEB model as given in Eq. \eqref{eq:2},

The term $k$ holds for the electron-electron exchange effects due to the interaction of the incident electron and the ejected electron.
\begin{equation}
    k = \frac{\ln t_{i}}{t_{i}+1}
\end{equation}
The reduced incident and orbital kinetic energy variables $t_{i},~u_{i}$ and the prefactor S are defined as,
\begin{equation}\label{eq:3}
S_i=4\pi a_0^2 N_i \left (\frac{R}{B_i}\right )^2,~u_i=\frac{U_i}{B_i},~t_i=\frac{E}{B_i}
\end{equation}
 The Bohr radius ($a_0 = 0.529\times10^{-10}m)$, Rydberg constant (${R = 13.6~eV}$), the orbital binding energy (B), the orbital kinetic energy (U), the orbital occupation number (N), the variable 
 (E or T ) can be taken as the kinetic energy of the incident electron.

The BEB model has gone through several modifications to calculate the PICS of the dissociating fragments from the parent molecule; significant contributions in this area include the semi-empirical works of Hamilton et al. \cite{hamilton2017calculated} the theoretical approach of Karl K. Irikura,\cite{irikura2017partial} also the mass spectrum dependence method (MSD) of Huber et al. \cite{huber2019total} and Graves et al. \cite{graves2022calculated}.
In this work, we calculate the PICS using the modified BEB (mBEB) formalism adopted from the works of Hamilton et al.  In our previous work \cite{shanmugasundaram2024electron} we have used the mBEB and the MSD models to calculate the PICS of various fragments appearing from 2-butanol in which we had a good agreement with the measured PICS. 
Although the above works have predicted the PICS due to electron impact, the area of fragmentation due to positron impact also has significant importance and is not explored as extensively as electron impact. 

The BEB model for positron impact has been deduced from the original BEB model by Fedus and Karwasaz \cite{fedus2019binary} and Franz et al. \cite{franz2021binary}. During positron impact ionization, the exit channel has a molecular ion and electron-positron annihilation.
Hence, there is no existence of electron-electron exchange effects, which makes $k = 0$. The collision of positrons at low incident energies induces positronium formation (Ps), which is not accounted for in the BEB model. Therefore, the process is termed direct ionization or direct positron impact ionization (DI). To estimate the total ionization cross sections (TICS) due to positron impact we should include the positronium formation cross sections $(\sigma{[\rm Ps]})$ in addition to the direct ionization,
\begin{equation}
    \sigma^{e^+}_{\rm TICS}(E) = \sigma{[\rm Ps]} + \sigma{[\rm DI]}
\end{equation} The estimation of the Ps cross sections is not our aim in this present work. So far there exist four BEB type models to calculate the DI cross sections ($\sigma{[\rm DI]}$). The BEB-0 and BEB-W models are introduced by Fedus and Karwasaz \cite{fedus2019binary} and the BEB-A and BEB-B models were presented by Franz et al \cite{franz2021binary}.
We will explore the BEB type models below here,
\subsection{BEB-0 and BEB-W} The simple BEB model for calculating the DICS is the BEB-0 model, where there are no scaling terms or any threshold laws were not implemented and yet calculates the cross sections accurately,
\begin{equation}
     \sigma_i^{BEB-0}(E) =\alpha \left [ \frac{ \ln t_i}{2}\left ( 1 - \frac{1}{t_i^{2}}\right) +  \left ( 1 - \frac{1}{t_{i}} \right )  \right ]
\end{equation}
The nature of the acceleration is not the same as electron impact in positron ionization, due to the repulsive character of the positron-nucleus interaction.
The term $k=0$ is applicable to the BEB model for positron impact, indicating the absence of electron-electron exchange interactions at shorter distances. As a result, the target electron cloud becomes distorted by the incoming positron.
This effect is significant because it gives rise to an attractive polarization potential, which enables the positron to overcome the nuclear repulsive potential. Consequently, this can lead to the formation of positronium (Ps) at low energies.

This leads to considering that the $u+1$ term holds for the acceleration of the incident positron due to attractive polarization. Fedus and Karwasaz found that in the threshold region the acceleration term is weak to reproduce the experimental profiles. In order to account for strong polarization effects and Ps formation at low energies, a new term has been added within the Burgess denominator in $\alpha$; such adaptation makes $\alpha \rightarrow \alpha^*$. 
\begin{equation}
    \alpha^* = \frac{\rm S_i}{(t_i+u_i+1+\rm X)}
\end{equation}
Now, depending on the approximation we imply on the factor X, the name of the model varies. Applying the Wannier-type threshold law to X makes the BEB-W model,\cite{fedus2019binary} incorporating the Jansen threshold makes it the BEB-A model.\cite{franz2021binary}

The BEB-W model is defined as, 
\begin{equation}\label{eq:beb-w}
      \sigma_i^{BEB-W}(E) = \alpha^*\left [\frac{\ln t_i}{2}\left ( 1 - \frac{1}{t_i^{2}} \right ) + 1 -\frac{1}{t_i} \right ]
\end{equation}
Here the factor X attributes for the Wannier type threshold law proposed by Klar \cite{Klar1981} from which the exponent 1.65 is deduced.
\begin{equation}
    \rm X= \frac{C}{(t_i - 1)^{1.65}}
\end{equation}
The value of constant C is not known as the Wannier threshold law only gives the proportionality of the cross section to the energy of the positron beyond the ionization threshold energy. Fixing the value of C to be unity reproduces the cross section profiles. Flipping the value of C to 0 gives us the BEB-0 model (the 0 in the superscript denotes C = 0 in BEB-W, hence the name BEB-0).

\subsection{BEB-A and BEB-B}
The BEB-W model in Eq. \eqref{eq:beb-w} has been further improved by Franz et al. \cite{franz2021binary} the factor X in the term $\alpha^*$ has been made to obey Jansen's threshold law.
\begin{equation}
        \rm X^A= \frac{C}{(t_i - 1)^{a-1}\exp{(-\beta_i\sqrt{t_i-1})}}
\end{equation}
The value of C is set to be unity as seen in the BEB-W model; the term $a = 2.540$, and the weighting factor $\beta_i$ is defined as,
\begin{equation}
    \beta_i = 0.489\sqrt{B_i/2R}
\end{equation}
This term depends on the orbital binding energy (B$_i$) of the molecule and follows Jansen's threshold law. Substituting $\rm X^A$ instead of X in $\alpha^*$ gives us the BEB-A model.

In a similar approach, the BEB-B model is constructed.
\begin{equation}
    \sigma_i^{BEB-B}(E)= \alpha \left [ \frac{\ln t_i }{2} \left ( 1 - \frac{1}{t_i^{2}} \right ) + h_i (t_i) \chi  + g_i(t_i)\chi^{a}  \right ]
\end{equation}
The pre-factors ($g_i , h_i$) and $\chi$ are defined as,
\begin{equation}
   g_i = C\exp{(-\beta_i\sqrt{t_i-1})},~ h_i = 1 - g_i,~ \chi = \left ( 1 - \frac{1}{t_i}  \right )
\end{equation}
At high energies, the effect of $g_i$ is negligible, making the second term $h_i$ more dominant, giving rise to the higher magnitude of the cross sections.
All these models were very well studied by Franz et al. \cite{franz2021binary} and Vincent Graves \cite{graves2024beb} and in our previous work\cite{suriyaprasanth2023electron} we have also studied electron and positron scattering on various sulfur-based compounds, in which we have calculated the ionization cross sections using the various BEB-type models.

In the recent work of Vincent Graves \cite{graves2024beb} have given a simple program RAPIDCS (relative and absolute partial ionization and dissociation- cross sections) for calculating the several electron and positron impact cross sections such as the total ionization, partial ionization, average secondary electron energy distribution, and stopping cross sections.
\section{Partial ionization cross sections}\label{sec:pics}
 This section focuses on the prerequisites and methods to calculate the PICS. As mentioned above, in order to calculate the PICS we must understand the fragmentation patterns of the molecule, and their branching ratios are needed.

The electron impact mass spectrum (EIMS) can be used to calculate the branching ratios at an incident energy resulting in the contributions of the cations arising from the parent molecules. Once the mass spectrum is measured, then the appearance energy of the cations is measured by fitting the ion yield curves on the Wannier threshold law.\cite{amorim2022-2butanol-1}

For simple polyatomic molecules, the fragmentation patterns can be guessed in a combinatorial way, and the dissociation energy of the emerging fragment can be calculated easily. For complex molecules, each fragment may contain several fragmentation pathways; hence, there is a great dependence on experimental measurements for the appearance energies. In principle, the terms appearance energy and dissociation energy mean the same property. To avoid any ambiguity among the readers, from here on we would stick to \say{dissociation energy ($\varepsilon$)}. 
As discussed above, we will use the mBEB method and the MSD method proposed by Goswami et al. \cite{butanol-goswami2022electron} and Graves et al \cite{graves2022calculated} in this work. These models have been very well tested by us in our previous work for 2-butanol \cite{shanmugasundaram2024electron}. So in this work, as an addition, we would also calculate the positron PICS using the BEBx (x = 0, A, B, W) models.
Before diving into our implementation of the PICS using the mBEB model and MSD methods, we will have a some introduction to calculation of the branching ratios.
\subsection{Branching ratios}
Like we discussed above, the EIMS measurement provides the relative abundances of the cations due to electron impact at single energy, which is up to the user and the limitations of the apparatus. Typically, EIMS is performed at 70 eV or 100 eV. 
\begin{equation}\label{eq:exp_br}
    \Gamma_i\rm(E_{r}) = \dfrac{\rm R(E_{r})}{\rm T(E_{r})}
\end{equation}
Here, $\rm R(E_{r})$ and $\rm T(E_{r})$ are the relative ion intensity and total ion intensity of the cations measured at the mass spectrum.
Although the EIMS data provide the BR at a single energy, branching ratios are incident kinetic energy dependent. Hence, the mass spectrum dependence method is being used to make the BR energy dependent.

\subsection{MSD method}
The BR from the EIMS data $\rm \Gamma_i(\rm E_r)$ is then scaled to reproduce the energy-dependent BR.
\begin{equation}\label{eq:17}
   \Gamma_i^{\rm MSD}(\rm E) =  \begin{cases}
0 & \text{ if } \rm E<\varepsilon \\ 
 \Gamma_i(\rm E_r)\left [ 1 - \left ( \dfrac{\varepsilon}{E} \right )^\mathit{z} \right ]& \text{ if } \rm E\geq \varepsilon
\end{cases}
\end{equation}
The control parameter $z$ is set to $1.5 \pm 0.2$ by Janev and Rieter,\cite{janev2004collision} further discussion involvinig the MSD method can be found in various sources.\cite{graves2022calculated,shanmugasundaram2024electron,graves2024beb} 

This is quite simple and does not require any scaling terms like the mBEB model, since the MSD BR itself is scaled with respect to the incident energy. The PICS calculated using MSD method is defined as,
\begin{equation}\label{eq:msd_pics}
    \sigma^{\rm PICS}_i = \Gamma_i^{\rm MSD}(\rm E) \times \sigma_i^{BEB}(E)
\end{equation}

\subsection{mBEB}
As discussed earlier, the mBEB model is a variant deduced from the BEB model by Hamilton et al. \cite{hamilton2017calculated} and further improvised by Goswami et al. \cite{goswami2021electron}. The key features of the mBEB model are that the dissociated cations of the parent molecule can be identified with respect to their dissociation energies ($\varepsilon$). The approach we take is by adding a correction term $(\delta)$ to the orbital binding energies of the parent molecule, which is defined as the difference between the dissociation of the cation and the ionization energy of the parent molecule, in short $\delta = |\text{IE} - \varepsilon|$. Such correction respectively to each of the cations alters the energy of the highest occupied molecular orbital to be the cation's $(\varepsilon)$. We can assume the modified binding energy as B', now we just replace the B $\rightarrow$ B' in the equation. \eqref{eq:3}, the S, $k$, $u_i, ~\rm and~ t_i \rightarrow S'$, $k'$, $u_i'$ and $t_i$', as the factor $\alpha$ holds all these values of S, $u_i, ~\rm and~ t_i$, which then becomes $\alpha'$
\begin{equation}\label{eq:u_mBEB}
   \sigma_i^{mBEB}~(E) =\alpha' \left [ \frac{ \ln {t_i'}}{2}\left ( 1 - \frac{1}{t_i'^{2}}\right) +  \left ( 1 - \frac{1}{t_{i}'} \right )- k'  \right ]
\end{equation}
Although such simple modification instigates the condition that the cross sections arise from the dissociation energy $\varepsilon$, this does not give us the correct cross sections we can call it as the \say{unscaled mBEB CS}. So, we also add a scaling factor $\Upsilon_i (\rm E_r)$, which reduces the magnitude of the cross section. This matches with the relative contribution of the cation with respect to its branching ratio from the EIMS. The scaling factor $\Upsilon_i$, is calculated from the ratio of the experimental BR $(\Gamma_i)$ and the theoretical BR $(\Gamma_i^T)$ as shown below,
\begin{equation}
    \Upsilon_i (E_r)  = \frac{\Gamma_i(\rm E_r) }{\Gamma_i^{T}(\rm E_r)}
\end{equation}
It has been earlier described about the experimental BR in Eq.\eqref{eq:exp_br}. To calculate the theoretical BR we have to take the ratio of the unscaled mBEB CS with the total ionization cross sections of the parent molecule. 
\begin{equation}\label{eq:11}
        \Gamma_i^{T}(\rm E_r)  = \frac{\sigma_i^{mBEB-B}(\rm E_r) }{\sigma_i^{BEB}(\rm E_r)}
    \end{equation}
To avoid any ambiguity in the cross sections, it is advised to know the reference energy in which the mass spectrum is measured and then do other calculations for the same energy. Once the scaling factor is obtained we can just calculate the mBEB PICS using,
\begin{equation}\label{eq:mbeb_pics}
    \sigma^{\rm PICS}_i (\rm E) = \Upsilon_i \times \sigma^{mBEB}(\rm E)
\end{equation}
From the Eq. \eqref{eq:msd_pics}, and Eq. \eqref{eq:mbeb_pics}, the electron impact partial ionization cross sections can be calculated. 
In order to calculate the positron impact partial ionization cross sections using the mBEB model, we are replacing the BEB model in Eq. \eqref{eq:u_mBEB} with the BEB-0 or BEB-W, to generalize this approach we call it mBEBx.

In a similar approach to calculate the positron PICS using the MSD method, we are using the positron TICS calculated using the BEB-0 or BEB-W in Eq. \eqref{eq:msd_pics}, this allows us to obtain the positron PICS using the MSDx model. In all the naming conventions, x is $x = 0,A,B,W$. For more details, one can look at the article by Vincent Graves where they have mentioned more about the PICS calculated using the BEBx models which are also integrated into the RAPIDCS program\cite{graves2024beb}. As suggested by Franz et al. \cite{franz2021binary}, the BEB-A and BEB-B models are best suited for nonpolar molecules, whereas the BEB-0 and BEB-W models are best suited for polar molecules. In the context of this manuscript, isobutanol is a polar molecule with a dipole moment of 1.640 Debye,\cite{nelson1967selected,johnson1999nist} Therefore, we stick to calculating and presenting the PICS and TICS of the positrons with BEB-0 and BEB-W models.	

\section{Results and Discussions}\label{sec:rd}

Orbital binding and kinetic energies were calculated using Gaussian-16 \cite{frisch2016gaussian} quantum chemistry software for the stable geometry of isobutanol. Geometry optimization was performed using density functional theory (DFT) with the $\omega \rm B97XD$ functional and the aug-cc-pVTZ basis set. The energy calculation was then performed at the Hartree-Fock (HF) approximation using the same basis set. The highest occupied molecular orbital (HOMO) binding energy was calculated to be 9.79 eV in DFT calculations, 11.94 eV in HF calculations and 13.52 eV in the QCxMS2 calculations using the GFN2-xTB level of theory\cite{bannwarth2019gfn2}.
Table. \ref{tab:ip}, contains the ionization potential calculated using the HF, DFT and QCxMS2 methods for isobutanol along with the comparison of the ionization potential of its isomers from the literature \cite{shanmugasundaram2024electron, bowen1984low, holmes1984heats, holmes1991ionization, linstrom2001nist}
The calculated binding energy (B) and kinetic energies (U) of the molecular orbitals using the HF method are shown in Table \ref{tab:mo_values}.

As we have mentioned earlier, the branching ratios are an important factor in calculating the PICS. In this work, we used quantum chemical mass spectrometry (QCxMS2)\cite{gorges2025qcxms2} to calculate the electron impact mass spectrum at incident energy of 70 eV. This is a very useful tool that provides a lot of information on the fragmentation of the molecule. In this paper, we just explore their synthetic mass spectrum from which we obtain the branching ratios.

\begin{table*}[h]
    \centering
    \caption{Ionization energy of isobutanol and its isomers: calculated HOMO and literature values.}
    \begin{tabular}{l c c}
        \toprule
        Molecule & HOMO (eV) & Literature (eV) \\ 
        \midrule
        isobutanol & 9.79, 11.94, 13.52 & $10.11\pm0.07$ \cite{bowen1984low}, $10.12\pm0.04$ \cite{holmes1984heats} \\  
        1-butanol & -- & $10.64\pm0.07$ \cite{bowen1984low}, $10.10\pm0.05$ \cite{holmes1991ionization}, $9.99\pm0.05$ \cite{linstrom2001nist} \\  
        2-butanol & 9.725 \cite{shanmugasundaram2024electron}, 11.790 \cite{shanmugasundaram2024electron} & $9.88\pm0.07$ \cite{bowen1984low}, $9.88\pm0.03$ \cite{holmes1984heats} \\  
        \bottomrule
    \end{tabular}
    \label{tab:ip}
\end{table*}

\begin{figure*}
    \centering
    \includegraphics[width=\linewidth]{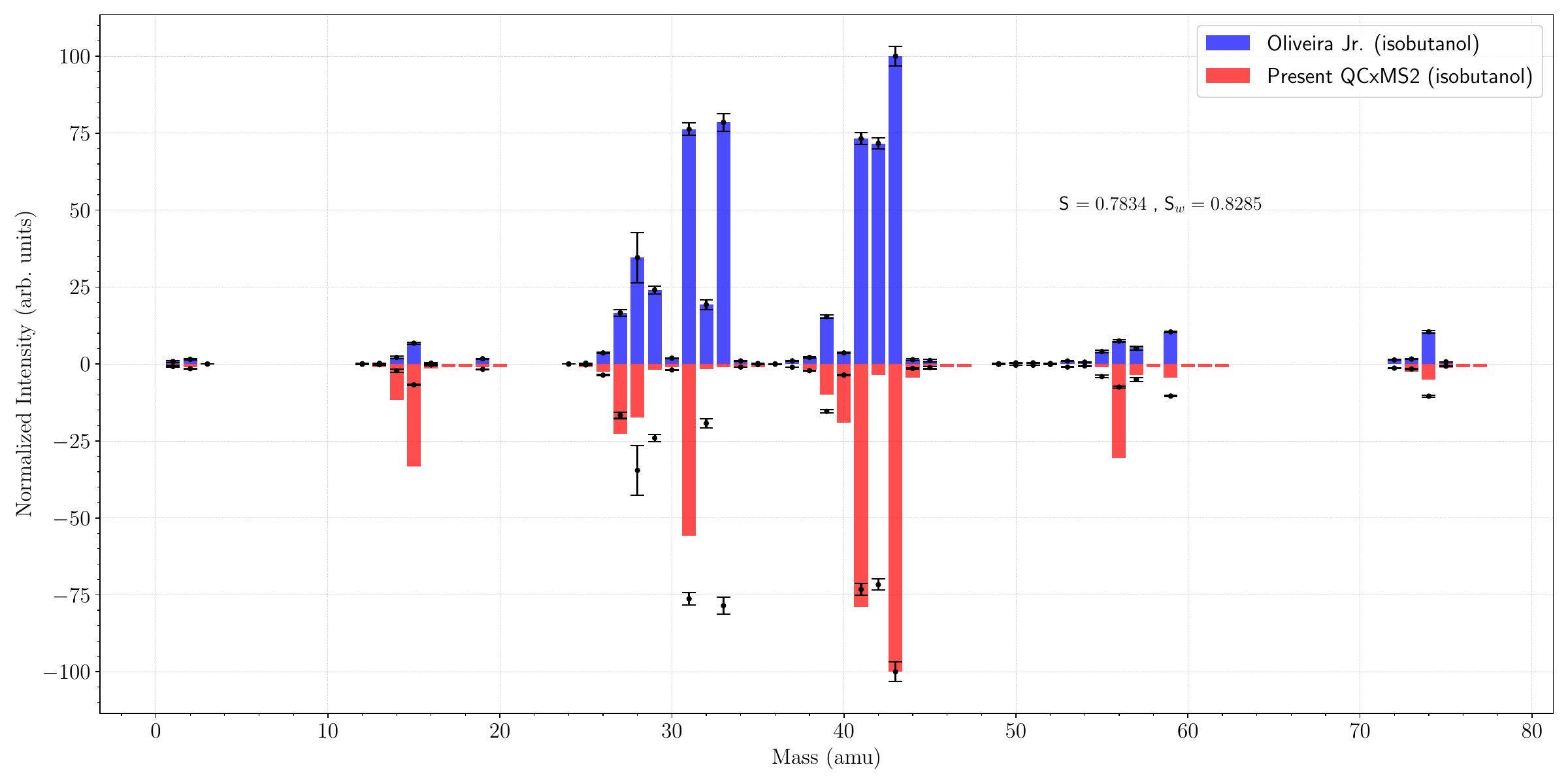}
    \caption{Comparision of electron impact mass spectrum using the QCxMS2 suite\cite{gorges2025qcxms2} with the measurements of Oliveira et al.\cite{oliveira2024ionic}  with uncertanity.}
    \label{fig:isobutanol_EIMS}
\end{figure*}

In figure \ref{fig:isobutanol_EIMS}, we have shown the comparison of the EIMS measurement of Oliveira et al.\cite{oliveira2024ionic} and the present calculations of EIMS using QCxMS2.  EIMS quantum chemistry calculations were performed at the GFN2 level\cite{bannwarth2019gfn2} for geometry optimizations and determination of the ionization potential using the $\omega b97x3c$ functional from ORCA\cite{RN204} for transition state calculations. From figure \ref{fig:isobutanol_EIMS}, we can see that the cosine score (S) and weighted cosine score ($S_w$) are 0.7834 and 0.8285. This tells us that there is good agreement between the EIMS measurements and the QCxMS2 calculations. The experimental uncertainty has also been included from the EIMS measurements, they have also been inverted and plotted on the $ -y$ axis to see how the QCxMS2 data compare with the experimental data. The present relative intensities (RI) of some cations are shown in the table. \ref{tab:all_eims} along with the experimental EIMS measurements \cite{oliveira2024ionic}. The branching ratios calculated from the QCxMS calculations are shown in table \ref{tab:e_pics_br}.  A more detailed study about the fragmentation of a molecule by electron impact will be presented in our future work.

\begin{table}[h]
    \centering
    \caption{The orbital binding energies (B) and orbital kinetic energies (U) are presented, calculated using the RHF/aug-cc-pVTZ method. The occupation number (N) is 2, and the dominant atomic orbital (AO) constituting the molecular orbital is shown.}
    \renewcommand{\arraystretch}{1.2}
    \begin{tabular}{rrrl}
        \toprule
        MO  & B (eV)  & U (eV)  & AO \\
        \midrule
        1A  & $559.321$ & $794.260$ & O $1s$   \\
        2A  & $306.664$ & $436.081$ & C $1s$   \\
        3A  & $305.346$ & $435.940$ & C $1s$   \\
        4A  & $305.085$ & $435.901$ & C $1s$   \\
        5A  & $304.839$ & $435.894$ & C $1s$   \\
        6A  & $37.010$  & $68.634$  & O $2s$   \\
        7A  & $29.363$  & $36.444$  & C $2s$   \\
        8A  & $25.241$  & $34.742$  & C $2s$   \\
        9A  & $24.856$  & $40.641$  & C $2s$   \\
        10A & $20.952$  & $34.273$  & C $2s$   \\
        11A & $18.920$  & $45.173$  & O $3p_x$ \\
        12A & $17.714$  & $29.424$  & C $3p_y$ \\
        13A & $16.529$  & $34.279$  & O $3p_z$ \\
        14A & $15.977$  & $31.087$  & C $3p_y$ \\
        15A & $14.747$  & $32.324$  & C $3p_x$ \\
        16A & $14.529$  & $41.871$  & O $3p_y$ \\
        17A & $14.369$  & $32.122$  & C $3p_z$ \\
        18A & $12.956$  & $38.161$  & C $4s$   \\
        19A & $12.844$  & $34.749$  & C $3p_x$ \\
        20A & $12.413$  & $36.793$  & C $4s$   \\
        21A & $11.944$  & $47.643$  & C $4p_z$ \\
        \bottomrule
    \end{tabular}
    \label{tab:mo_values}
\end{table}

\begin{figure*}[ht!]
    \centering
    \includegraphics[width=\linewidth]{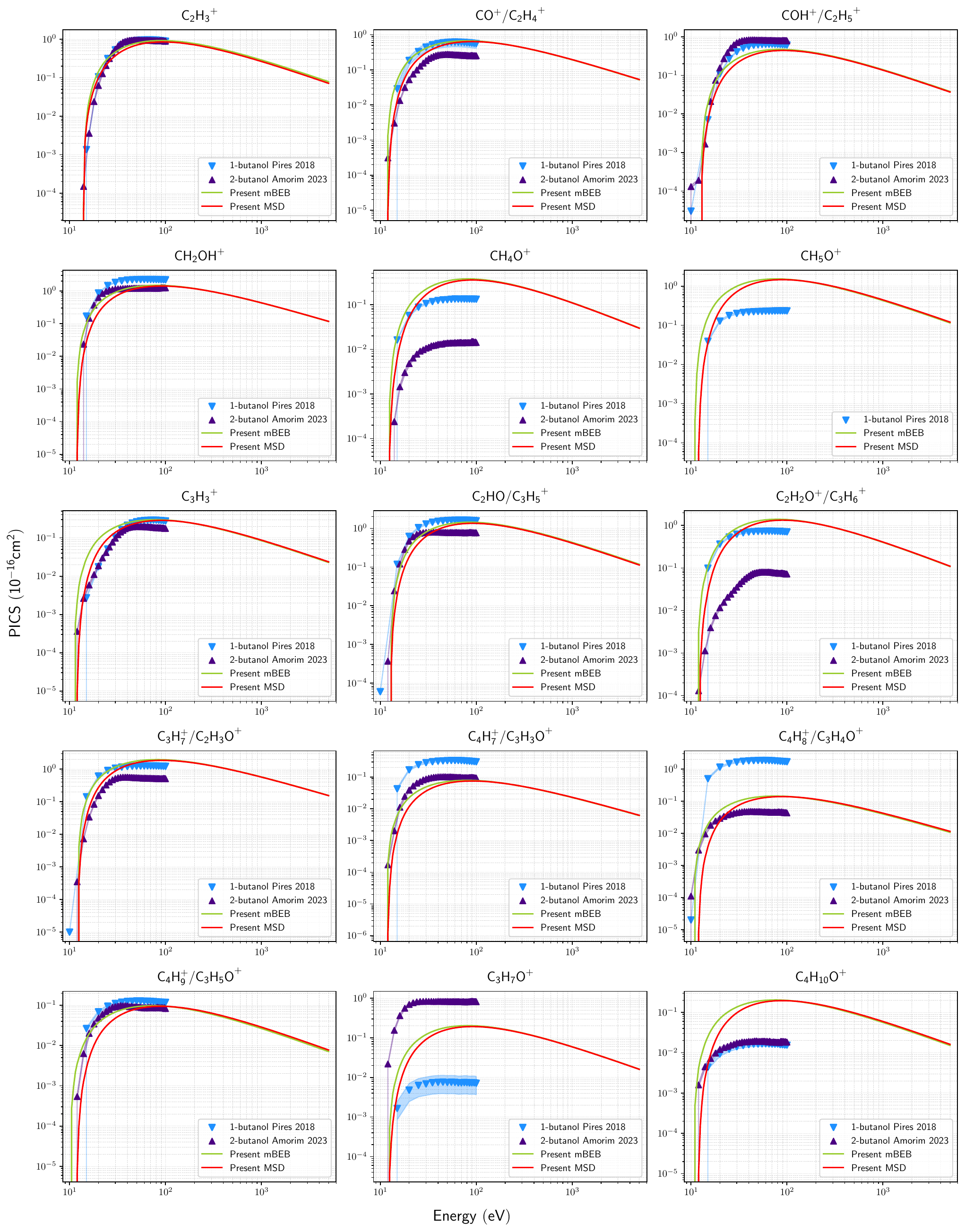}
    \caption{Our calculated partial ionization cross sections of isobutanol using mBEB (solid green line) and MSD method (soild red line) are compared with the experimental measurements of 1-butanol\cite{pires2018electron}(inverted triangle) and 2-butanol\cite{amorim20232-butanol-2}(upright triangle).}
    \label{pics_all_exp}
\end{figure*}

 Figure \ref{pics_all_exp}, shows the comparison of the calculated PICS for isobutanol with the experimental data of 1-butanol\cite{pires2018electron} and 2- butanol\cite{amorim2021absolute} for electron impact. Since the PICS of isobutanol is unavailable in the literature, we are comparing our PICS with the measurements of isobutanol isomers, such as 1-butanol and 2-butanol. We have used Olivera et al. EIMS data \cite{oliveira2024ionic} in our calculations of the branching ratio and PICS, as they have given more detailed information of most cationic fragments compared to the NIST Web book\cite{linstrom2001nist} and the spectral database of organic compounds (SDBS)\cite{SDBS}. The table \ref{tab:e_pics_br} contains the dissociation energy of the cations measured ($\varepsilon$) by Olivera et al. \cite{oliveira2024ionic} and the branching ratios $(\Gamma_i)$ calculated from their electron impact mass spectrum data. In the same table, we have compared the calculated electron impact PICS of isobutanol with the PICS of 1-butanol and 2-butanol available in the literature.\cite{pires2018electron,butanol-goswami2022electron,amorim20232-butanol-2,shanmugasundaram2024electron} As shown in our previous study on 2-butanol,\cite{shanmugasundaram2024electron} the calculated PICS has a good comparison with the experimental cross sections for most of the cations.  The PICS of $\mathsf{C_2H_3}^+$ agrees very well with the measurements of 1-butanol and 2-butanol; the contribution of the fragment to isobutanol, 1-butanol, and 2-butanol is 43.63, 43.0, and 37.35 percentages. A similar trend can be seen for the cation $\mathsf{CO/C_2H_4}^+$, where abundances are 24.05, 27.17 and 10.43 percentages for the cation in the measurements of isobutanol, 1-butanol, and 2-butanol. These trends can be observed in cations such as $\mathsf{CH_2OH}^+$ and $\mathsf{C_3H_3}^+$. For all other cations of isobutanol, there is a large difference in the PICS comparison of similar fragments of 1-butanol and 2-butanol due to the difference in cation abundances.
The detailed cationic abundances of isobutanol, 1-butanol, and 2-butanol are shown in table \ref{tab:all_eims}.
It has been verified that the cation identified as base peak (with RI = 100) will have a larger PICS compared to the cross sections of other cations in their set $\text{(i.e., for } \sigma_{\max}(\mathsf{CH_5O^+}) \text{ is } \approx 77.2\% \text{ of } 
\mathsf{C_3H_7}^+/\mathsf{C_2H_3O^+} \text{ in isobutanol)}$, this is true for all and can be verified. All the PICS contributions of the other cations with respect to the base cation are the same as their relative abundance. These assumptions have been observed from the PICS compared in figure \ref{pics_all_exp}, and the relative abundances seen from the table \ref{tab:all_eims}.
\begin{table*}[h!]
    \centering
    \caption{Relative abundances of cations of isobutanol compared with the reported abundances of 1-butanol and 2-butanol cations. The cation \ce{C2H5O} was found to be 100 at the 2-butanol EIMS measurements \cite{amorim2022-2butanol-1}.}
    \setlength{\tabcolsep}{3pt}  
    \renewcommand{\arraystretch}{1.2}  
    \begin{tabular}{c l l l l l}
        \toprule
       \multirow{3}{*}{$m/z$}&\multirow{3}{*}{Cation}& &\multicolumn{2}{c}{Relative abundances}\\
       \cmidrule(lr){3-6}
         &  & \multicolumn{2}{c}{isobutanol} & \multirow{2}{*}{1-butanol\cite{pires2018electron}} & \multirow{2}{*}{2-butanol\cite{amorim2022-2butanol-1}} \\
        \cmidrule(lr){3-4}
        & & Present & Oliveira \cite{oliveira2024ionic} &  &  \\
        \midrule
        27  & \ce{C2H3}          & 22.57  & 46.63  & 43.00  & 37.35  \\
        28  & \ce{CO}/\ce{C2H4}  & 17.28  & 34.56  & 27.17  & 10.43  \\
        29  & \ce{COH}/\ce{C2H5} & 1.78   & 24.05  & 26.68  & 32.22  \\
        31  & \ce{CH2OH}         & 55.76  & 76.28  & 100.00 & 48.82  \\
        32  & \ce{CH4O}          & 1.62   & 19.80  & 5.94   & 0.57   \\
        33  & \ce{CH5O}          & 1.22   & 78.51  & 10.28  & 0.17   \\
        39  & \ce{C3H3}          & 9.84   & 15.40  & 12.98  & 7.57   \\
        41  & \ce{C3H5}/\ce{C2HO} & 78.87  & 73.24  & 72.78  & 30.78  \\
        42  & \ce{C3H6}/\ce{C2H2O} & 3.62  & 71.68  & 32.60  & 3.15   \\
        43  & \ce{C3H7}/\ce{C2H3O} & 100.00 & 100.00 & 55.62  & 20.76  \\
        55  & \ce{C4H7}/\ce{C3H3O} & 1.09   & 4.06   & 14.48  & 1.23   \\
        56  & \ce{C4H8}/\ce{C3H4O} & 30.56  & 7.48   & 89.93  & 1.81   \\
        57  & \ce{C4H9}/\ce{C3H5O} & 3.56   & 5.05   & 5.55   & 3.51   \\
        59  & \ce{C3H7O}         & 1.08   & 10.42  & 0.33   & 32.65  \\
        74  & \ce{C4H10O}        & 5.11   & 10.46  & 0.73   & 0.73   \\
        \bottomrule
    \end{tabular}
    \label{tab:all_eims}
\end{table*}

\begin{figure*}[ht!]
    \centering
    \includegraphics[width=\linewidth]{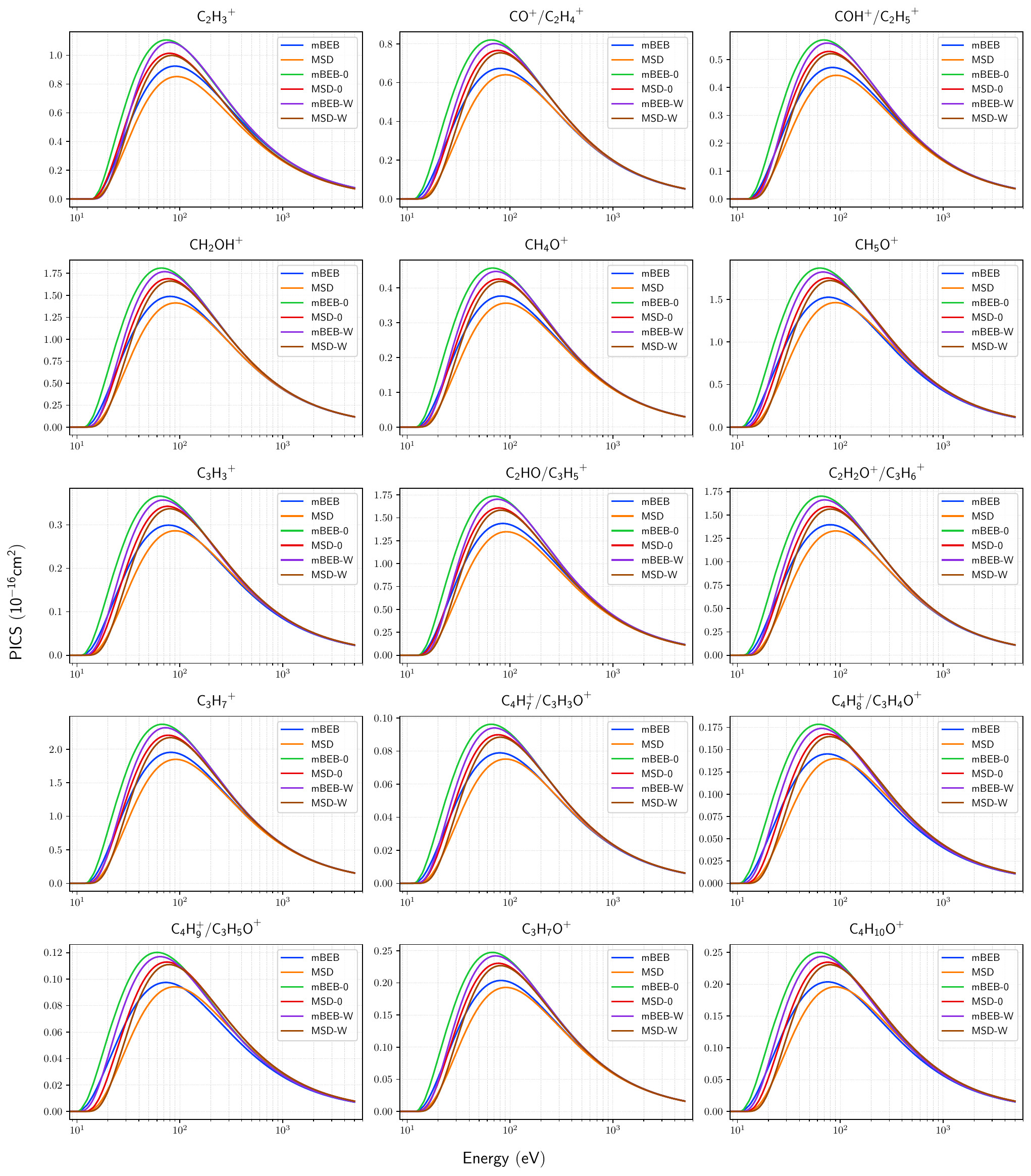}
    \caption{Partial ionization cross sections of isobutanol for electron and positron impact}
    \label{e_pics_isobutanol}
\end{figure*}

In the absence of any other comparison for PICS of isobutanol and to gain more insight, here we present electron- and positron-impact PICS, calculated using the mBEB and the MSD methods, shown in the figure. \ref{e_pics_isobutanol}.  Here, in Figure \ref{e_pics_isobutanol}, we have presented the PICS of $\mathsf{C_2H_3}^+$, $\mathsf{CO}^+/\mathsf{C_2H_4}^+$, $\mathsf{COH}^+/\mathsf{C_2H_5}^+$, $\mathsf{CH_2OH}^+$, $\mathsf{CH_4O}^+$, $\mathsf{CH_5O}^+$, $\mathsf{C_3H_3}^+$, $\mathsf{C_2HO}^+/ \mathsf{C_3H_5}^+$, $\mathsf{C_2H_2O}^+/ \mathsf{C_3H_6}^+$, $\mathsf{C_3H_7}^+$, $\mathsf{C_4H_7}^+/ \mathsf{C_3H_3O}^+$, $\mathsf{C_4H_8}^+ /\mathsf{C_3H_4O}^+$, $\mathsf{C_4H_9}^+ /\mathsf{C_3H_5O}^+$, $\mathsf{C_3H_7O}^+$,
$\mathsf{C_4H_{10}O}^+$. From the EIMS measurements it can be seen that isobutanol is more stable than its isomers (1-butanol and 2-butanol). Isobutanol has less fragments (45 cations), which is significantly less than 1-butanol and 2-butanol. This is one point that makes isobutanol important, as it has fewer by-products, making it a good alternative to conventional petroleum-based products. Bringing us to the PICS, the trend observed here is nominal; as usual, the positron PICS has a higher magnitude compared to the for all fragments of isobutanol.

The $\mathsf{C_3H_7}$ cation is the most stable with a mass of 43 amu, observed in the mass spectrum. This cation has the greatest magnitude, as expected. Including the base fragment $\mathsf{C_3H_7}^+  (43)$, other cations such as $ \mathsf{C_2H_3}^+(27) $, $ \mathsf{CH_2OH}^+(31) $, $ \mathsf{CH_5O}^+(33) $, $ \mathsf{C_2HO}^+/\mathsf{C_3H_5}^+(41) $, $ \mathsf{C_2H_2O}^+/\mathsf{C_3H_6}^+(42) $ contribute to 50 \% of the mass spectrum. This also implies summing up their PICS yields 50 \% of the TICS. This is true for both electron- and positron-impact-ionization cross sections. In table \ref{tab:e_pics_br}, we have shown the comparison of PICS of the cationic fragments which we discussed. All 15 of these cations have also been detected in 1-butanol and 2-butanols, and they have a different relative intensity. Most of the cations in this set except $\mathsf{CH_5O} ^+(33)$ have measured PICS in 2-butanol. Since $\mathsf{CH_5O}^+$ only had an abundance of 0.17\%, it must be difficult to measure its PICS in an experiment. The mBEB data presented by Goswami et al. for 1-butanol\cite{butanol-goswami2022electron} and the MSD data for 2-butanol from our previous work\cite{shanmugasundaram2024electron} have been compared with the absolute experimental PICS measurements. There would be some disagreements at the cross sections in the low-energy or near-threshold region, as the BEB model works very well for high-energy. Studies on 1-butanol and 2-butanol \cite{butanol-goswami2022electron,shanmugasundaram2024electron} show that PICS agrees well with experimental measurements where the incident energies are greater than 35 eV.
The difference in cross-sectional peak positions for cations observed in the experiment indicates that the energy dependence of each reaction channel varies from that of the TICS. This shows that various dissociation mechanisms influence each fragmentation pathway. These differences are visible for both electron- and positron-impact PICS, in the low-energy region the PICS calculated using the mBEBx and MSDx models are clearly distinguishable., in all these models the Burgees denominator ($\alpha^*$) and the scaling factor X play a vital role, as mentioned in sec \ref{sec:level2}. The BEB model of the positron (BEB-0, BEB-W) has been well investigated in the literature \cite{fedus2019binary,franz2021binary,suriyaprasanth2023electron} and for PICS using the MSD method the BEB-0 and BEB-W\cite{graves2024beb} models are tested. So, the overall trend for electron impact and positron impact that is seen here is valid.

\begin{figure}[ht]
    \centering
    \includegraphics[width=0.5\textwidth]{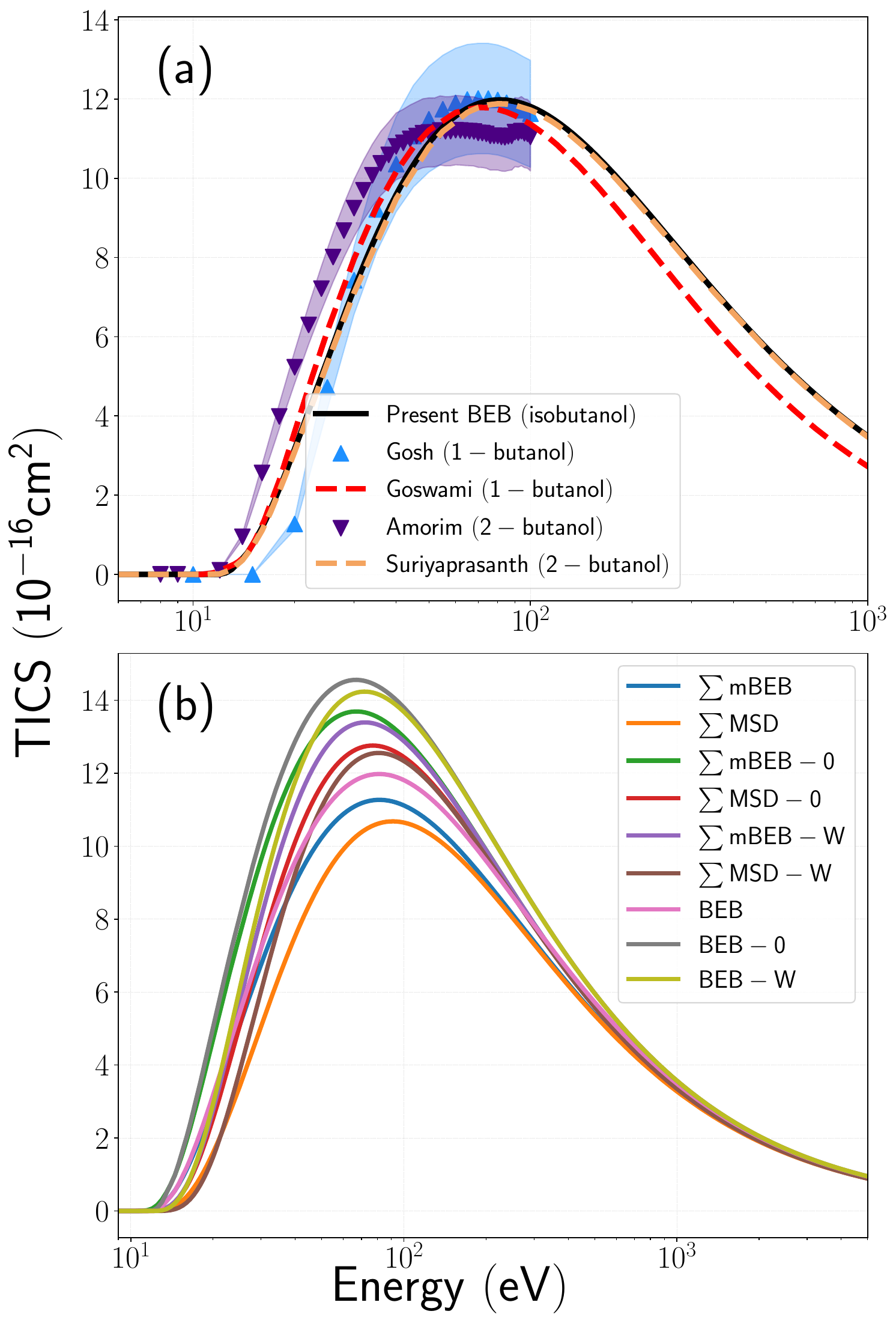}
     \caption{a) Comparison of TICS plots for isobutanol with 1-butanol\cite{ghosh2018electron,butanol-goswami2022electron} and 2-butanol\cite{ghosh2018electron,shanmugasundaram2024electron} TICS, b) The comparison of TICS due to electron impact and positron impact and the summation of the partial ionization cross sections calculated with various methods have been shown. }
    \label{fig:tics}
\end{figure}

In Figure. \ref{fig:tics}a, we have compared the TICS of isobutanol with their isomers for electron impact; it shows that the magnitudes of our current TICS of isobutanol are slightly higher than the BEB TICS of 1-butanol\cite{butanol-goswami2022electron} and 2-butanol.\cite{amorim20232-butanol-2,shanmugasundaram2024electron} However, the present data lie within the experimental uncertainty of these isomers. Since there were no data available for comparison for the positron TICS, we have plotted all the positron TICS together with electron impact TICS in the figure. \ref{fig:tics} along with the sum of the PICS of the mBEB and MSD methods. The positron impact TICS has a larger magnitude compared to the electron impact data in general.

\begin{table*}
 \centering
    \setlength{\tabcolsep}{3pt} 
    \caption{The branching ratios and electron impact partial cross sections compared with the literature data of 1-butanol\cite{pires2018electron} and 2-butanol\cite{amorim20232-butanol-2}.}  
    \begin{tabular}{lllllllllll}
    \toprule
    $m/z$ & Cation & $\varepsilon$ (eV)\cite{oliveira2024ionic} & \multicolumn{2}{c}{$\Gamma_i$} & \multicolumn{2}{c}{Present isobutanol} & \multicolumn{2}{c}{1-butanol} & \multicolumn{2}{c}{2-butanol} \\ 
    \multicolumn{4}{c}{} Oliveira \cite{oliveira2024ionic} & Present & $\sigma_{\rm MSD}^{\rm max}$(E) & $\sigma_{\rm mBEB}^{\rm max}$(E) & $\sigma_{\rm exp}^{\rm max} $(E) \cite{pires2018electron} & $\sigma_{\rm mBEB}^{\rm max}$(E)\cite{butanol-goswami2022electron} & $\sigma_{\rm exp}^{\rm max}$ (E)\cite{amorim20232-butanol-2} & $\sigma_{\rm MSD}^{\rm max}$ (E)\cite{shanmugasundaram2024electron} \\ 
    \midrule
    $27$ & $\rm C_2H_3$ & $13.99$ & $0.0760$ & 0.0518 & $0.852(93.5)$ & $0.925(90.0)$ & $0.965(70)$ & $0.946(80.0)$ & $0.964(48)$ & $0.265(90.0)$ \\
   $28$ & $\rm CO/C_2H_4$ & $11.64$ & $0.0563$ & 0.0397 & $0.641(90.5)$ & $0.674(80.0)$ & $0.616(60)$ & $0.596(75.0)$ & $0.279(50)$ & $0.287(90.0)$ \\
    $29$ & $\rm COH/C_2H_5$ & $12.6$ & $0.0392$ & 0.0041 & $0.444(92.0)$ & $0.472(84.0)$ & $0.644(70)$ & $0.629(75.0)$ & $0.834(44)$ & $0.332(90.0)$ \\
    $31$ & $\rm CH_2OH$ & $11.79$ & $0.1243$ &0.1281 & $1.414(91.0)$ & $1.488(80.5)$ & $2.245(65)$ & $2.306(70.0)$ & $1.271(90)$ & $0.317(90.0)$ \\
    $32$ & $\rm CH_4O$ & $12.19$ & $0.0314$ &0.0037 & $0.356(91.5)$ & $0.377(82.0)$ & $0.134(75)$ & $0.130(70.0)$ & $0.015(90)$ & $0.307(90.0)$ \\
    $33$ & $\rm CH_5O$ & $10.96$ & $0.1280$ & 0.0026& $1.463(90.0)$ & $1.526(77.0)$ & $0.235(95)$ & $0.225(70.0)$ & $-$ & $-$ \\
    $39$ & $\rm C_3H_3$ & $11.17$ & $0.0251$ & 0.0226& $1.463(90.0)$ & $0.300(78.0)$ & $0.290(75)$ & $0.282(70.0)$ & $0.198(48)$ & $0.359(100.0)$ \\
    $41$ & $\rm C_3H_5/C_2HO$ & $12.84$ & $0.1194$ & 0.1812& $1.348(92.0)$ & $1.439(85.0)$ & $1.633(70)$ & $1.593(70.0)$ & $0.801(34)$ & $0.304(90.0)$ \\
    $42$ & $\rm C_3H_6/C_2H_2O$ & $11.6$ & $0.1168$ & 0.0083& $1.330(90.5)$ & $1.397(79.5)$ & $0.733(65)$ & $0.714(70.0)$ & $0.081(56)$ & $0.338(90.0)$ \\
    $43$ & $\rm C_3H_7/C_2H_3O$ & $12.11$ & $0.1630$ & 0.2297& $1.850(91.0)$ & $1.954(82.0)$ & $1.248(70)$ & $1.218(70.0)$ & $0.566(40)$ & $0.291(90.0)$ \\
    $55$ & $\rm C_4H_7/C_3H_3O$ & $11.58$ & $0.0066$ & 0.0025& $0.075(90.5)$ & $0.079(79.5)$ & $0.333(60)$ & $0.321(70.0)$ & $0.102(46)$ & $0.304(90.0)$ \\
    $56$ & $\rm C_4H_8/C_3H_4O$ & $10.5$ & $0.0122$ & 0.0702& $0.140(89.5)$ & $0.145(75.0)$ & $1.873(55)$ & $1.794(70.0)$ & $0.048(42)$ & $0.237(90.0)$ \\
    $57$ & $\rm C_4H_9/C_3H_5O$ & $10.07$ & $0.0082$ &0.0082 & $0.095(89.0)$ & $0.098(73.0)$ & $0.127(55)$ & $0.122(70.0)$ & $0.097(34)$ & $0.304(90.0)$ \\
    $59$ & $\rm C_3H_7O$ & $12.06$ & $0.0170$ &0.0025 & $0.193(91.0)$ & $0.204(81.5)$ & $0.008(45)$ & $0.007(70.0)$ & $0.840(90)$ & $0.374(100.0)$ \\
    $74$ & $\rm C_4H_{10}O$ & $10.61$ & $0.0171$ & 0.0117& $0.195(89.5)$ & $0.203(75.5)$ & $0.017(60)$ & $0.016(70.0)$ & $0.020(48)$ & $0.250(90.0)$ \\
    \bottomrule
\end{tabular}
\label{tab:e_pics_br}

\end{table*}

\section{Conclusion}
The electron impact and positron impact PICS and TICS of isobutanol have been calculated. It is found that isobutanol is more stable in the $\mathsf{C_4-C_4}$ alcohol set. It is also proved that isobutanol is very much more stable than their isomers, such as 1-butanol or 2-butanol or n-butanol. Hence, this could be used in a study that seeks a viable alternative to current fossil fuels\cite{oliveira2024ionic}. 
The PICS resulting from electron and positron impact have been calculated and presented for the first time in the literature.
The orbital binding energies and orbital kinetic energies were also presented with the dominant atomic orbital contribution to the construction of the molecular orbital. The QCxMS2 program has been tested to calculate the synthetic electron impact mass spectrum; this also has a fairly good comparison with the experimental data. Although QCxMS2 did not calculate the correct intensity for some fragments, the predictions were in agreement with the experimental measurements for some prominent fragments; Refer to table \ref{tab:supp_a1}. A1. In future work, along with the TICS and PICS, we will perform an exclusive study for the fragmentation of a polyatomic molecule.

\subsection*{Acknowledgements}
The authors acknowledge the Science and Engineering Research Board (SERB), Department of Science and Technology (DST), Government of India (Grant No. SRG/2022/000394) for providing a computing facility.

\subsection*{Data availability}
All data generated or analyzed during this study will be available as supplementary material.

\subsection*{Author contributions statement}
\textbf{Suriyaprasanth Shanmugasundaram:} Conceptualization
(lead), Data curation (lead), Investigation (lead), Methodology(Supporting), Resources (Supporting), Validation (lead),Writing—original draft (lead), Writing—review \& editing (equal); \textbf{Dhanoj Gupta:} Conceptualization (equal), Data curation (equal), Investigation (equal), Methodology (lead), Resources (lead), Validation (equal), Writing original draft (equal), Writing—review \& editing (lead), Supervision (lead). All authors reviewed the manuscript.

\appendix

\section*{Appendix A. Supplementary information}

The relative abundances calculated from the QCxMS2 and the literature are shown in this table. \ref{tab:supp_a1}. A1.

\begin{table}[h]
    \centering
    \begin{tabular}{lccc}
        \toprule
        
        $m/z$ & Present & EIMS\cite{oliveira2024ionic} & Error\cite{oliveira2024ionic} \\
        &QCxMS2&&\\
        \midrule
        1   & $1.00$  & 0.81       & 0.31   \\
        2   & 1.00  & 1.52       & 0.15   \\
        3   & -      & 0.01       & 0.01   \\
        12  & -      & 0.1        & 0.04   \\
        13  & 1.00  & 0.24       & 0.04   \\
        14  & 11.68 & 2.15       & 0.49   \\
        15  & 33.37 & 6.76       & 0.29   \\
        16  & 1.36  & 0.29       & 0.23   \\
        17  & 1.00  & -          & -      \\
        18  & 1.00  & -          & -      \\
        19  & 1.00  & 1.71       & 0.12   \\
        20  & 1.00  & -          & -      \\
        24  & -      & 0.04       & 0.01   \\
        25  & 1.00  & 0.3        & 0.04   \\
        26  & 2.46  & 3.6        & 0.17   \\
        27  & 22.57 & 16.63      & 1.05   \\
        28  & 17.28 & 34.56      & 8.15   \\
        29  & 1.78  & 24.05      & 1.23   \\
        30  & 1.01  & 1.91       & 0.12   \\
        31  & 55.76 & 76.28      & 2.01   \\
        32  & 1.62  & 19.28      & 1.53   \\
        33  & 1.12  & 78.51      & 2.85   \\
        34  & 1.00  & 0.99       & 0.06   \\
        35  & 1.00  & 0.17       & 0.03   \\
        36  & -      & 0.06       & 0.03   \\
        37  & -      & 1.04       & 0.08   \\
        38  & 1.86  & 2.15       & 0.11   \\
        39  & 9.84  & 15.4       & 0.55   \\
        40  & 19.07 & 3.6        & 0.15   \\
        41  & 78.87 & 73.24      & 1.96   \\
        42  & 3.62  & 71.68      & 1.81   \\
        43  & 100.0 & 100.0      & 3.18   \\
        44  & 4.33  & 1.42       & 0.27   \\
        45  & 1.03  & 1.16       & 0.39   \\
        46  & 1.00  & -          & -      \\
        47  & 1.00  & -          & -      \\
        49  & -      & 0.12       & 0.02   \\
        50  & -      & 0.39       & 0.05   \\
        51  & -      & 0.39       & 0.06   \\
        52  & -      & 0.18       & 0.03   \\
        53  & -      & 0.99       & 0.09   \\
        54  & -      & 0.61       & 0.09   \\
        55  & 1.09  & 4.06       & 0.42   \\
        56  & 30.56 & 7.48       & 0.33   \\
        57  & 3.56 & 5.05       & 0.62   \\
        58  & 1.08  & -          & -      \\
        59  & 4.37  & 10.42      & 0.16   \\
        60  & 1.11  & -          & -      \\
        61  & 1.00  & -          & -      \\
        62  & 1.00  & -          & -      \\
        72  & -      & 1.33       & 0.09   \\
        73  & 2.61  & 1.62       & 0.15   \\
        74  & 5.11  & 10.46      & 0.35   \\
        75  & 1.18  & 0.71       & 0.06   \\
        76  & 1.01  & -          & -      \\
        77  & 1.00  & -          & -      \\
        \bottomrule
    \end{tabular}
    \caption{$\mathsf{\bf A1.}$ Comparison of calculated QCxMS2 data with measured electron ionization mass spectrometry (EIMS) data of Oliveira et al\cite{oliveira2024ionic}}
    \label{tab:supp_a1}
\end{table}


\bibliography{main}

\end{document}